\documentclass[12pt,compress,nosort]{JHEP3}
\usepackage{exscale,latexsym}
\usepackage[dvips]{graphicx}
\usepackage{axodraw}
\usepackage{cite}

\def\Year{\expandafter\eatPrefix\the\year}
\newcount\hours \newcount\minutes
\def\monthname{\ifcase\month\or
January\or February\or March\or April\or May\or June\or July\or
August\or September\or October\or November\or December\fi}
\def\shortmonthname{\ifcase\month\or
Jan\or Feb\or Mar\or Apr\or May\or Jun\or Jul\or Aug\or Sep\or
Oct\or Nov\or Dec\fi}

\def\TimeStamp{\hours\the\time\divide\hours by60%
\minutes -\the\time\divide\minutes by60\multiply\minutes by60%
\advance\minutes by\the\time%
${\rm \shortmonthname}\cdot   \if\day<10{}0\fi\the\day\cdot
\the\year \qquad\the\hours:\if\minutes<10{}0\fi\the\minutes$}

\newskip\humongous \humongous=0pt plus 1000pt minus 100pt
\def\caja{\mathsurround=0pt}
\def\eqalign#1{\,\vcenter{\openup1\jot \caja
       \ialign{\strut \hfil$\displaystyle{##}$&$
        \displaystyle{{}##}$\hfil\crcr#1\crcr}}\,}
\newif\ifdtup

\newcounter{eqnumber}[section]
\renewcommand{\theeqnumber}{\thesection.\arabic{eqnumber}}
\def\equn{\refstepcounter{eqnumber}
\eqno({\rm \theeqnumber}) }

\def\tree{{\rm tree}}

\newbox\charbox
\newbox\slabox
\def\s#1{{      
        \setbox\charbox=\hbox{$#1$}
        \setbox\slabox=\hbox{$/$}
        \dimen\charbox=\ht\slabox
        \advance\dimen\charbox by -\dp\slabox
        \advance\dimen\charbox by -\ht\charbox
        \advance\dimen\charbox by \dp\charbox
        \divide\dimen\charbox by 2
        \raise-\dimen\charbox\hbox to \wd\charbox{\hss/\hss}
        \llap{$#1$}
}}

\def\spa#1.#2{\left\langle#1\,#2\right\rangle}
\def\spb#1.#2{\left[#1\,#2\right]}
\def\spba#1.#2.#3{\left[#1|#2|#3\right\rangle}
\def\spab#1.#2.#3{\left\langle#1|#2|#3\right]}
\def\spaa#1.#2.#3{\left\langle#1|#2|#3\right\rangle}
\def\spbb#1.#2.#3{\left[#1|#2|#3\right]}
\def\lor#1.#2{\left(#1\,#2\right)}

\catcode`@=11  

\def\Slash#1{\slash\hskip -0.22 cm #1}

\def\la{\langle}
\def\ra{\rangle}

\def\lsl{\not{\hbox{\kern-2.3pt $\ell$}}}
\def\Psl{\not{\hbox{\kern-2.3pt $P$}}}
\def\ksl{\not{\hbox{\kern-2.3pt $k$}}}
\def\twosl{\not{\hbox{\kern-2.3pt $2$}}}
\def\fivesl{\not{\hbox{\kern-2.3pt $5$}}}

\def\spa#1.#2{\left\langle#1\,#2\right\rangle}
\def\spb#1.#2{\left[#1\,#2\right]}
\def\lor#1.#2{\left(#1\,#2\right)}

\def\sand#1.#2.#3{%
  \left\langle\smash{#1}{\vphantom1}\right|{#2}%
  \left|\smash{#3}{\vphantom1}\right\rangle}
\def\sandp#1.#2.#3{%
  \left\langle\smash{#1}{\vphantom1}^{-}\right|{#2}%
  \left|\smash{#3}{\vphantom1}^{+}\right\rangle}
\def\sandpp#1.#2.#3{%
  \left\langle\smash{#1}{\vphantom1}^{+}\right|{#2}%
  \left|\smash{#3}{\vphantom1}^{+}\right\rangle}
\def\sandmm#1.#2.#3{%
  \left\langle\smash{#1}{\vphantom1}^{-}\right|{#2}%
  \left|\smash{#3}{\vphantom1}^{-}\right\rangle}
\def\sandpm#1.#2.#3{%
  \left\langle\smash{#1}{\vphantom1}^{+}\right|{#2}%
  \left|\smash{#3}{\vphantom1}^{-}\right\rangle}
\def\sandmp#1.#2.#3{%
  \left\langle\smash{#1}{\vphantom1}^{-}\right|{#2}%
  \left|\smash{#3}{\vphantom1}^{+}\right\rangle}

\def\Atree{A^{\rm tree}}

\def\BRQ#1#2#3{\la #1|{#2}|#3\ra}
\def\BBRQ#1#2#3{[ #1|{#2}|#3\ra}

\def\small{}

\title{MHV-Vertices for Gravity Amplitudes}

\author{N.~E.~J.~Bjerrum-Bohr${}^1$, David~C.~Dunbar${}^1$,
Harald Ita${}^1$,
Warren~B.~Perkins${}^1$ and Kasper Risager${}^2$
\\
${}^1$ Department of Physics,
University of Wales Swansea, \\
\,\, Swansea, SA2 8PP, UK
\\
${}^2$
Niels Bohr Institute,
Blegdamsvej 17, DK-2100, 
Copenhagen,
Denmark}

\preprint{hep-th/0509016\\ SWAT-05-439}

\abstract{
We obtain a CSW-style formalism for calculating graviton scattering amplitudes and prove its
validity through the use of a special type of
 BCFW-like parameter shift.
The procedure is illustrated with explicit examples.
}

\keywords{Models of Quantum Gravity, Supergravity Models}

\begin{document}
\section{Introduction}

In a helicity formalism the simplest Yang-Mills amplitudes are the
MHV amplitudes where  precisely two external gluons have negative
helicity and the remaining legs all have positive helicity.
If legs  $j$ and $k$ have negative helicity, the colour-ordered~\cite{TreeColour} partial
amplitude takes the form~\cite{ParkeTaylor},
$$
\eqalign{
  \Atree_n(1^+,\ldots,j^-,\ldots,k^-,
                \ldots,n^+)\,
&\;=\;\ i\, { {\spa{j}.{k}}^4 \over \spa1.2\spa2.3\cdots\spa{n}.1 }\,.
  \cr}
\equn\label{ParkeTaylor}
$$
We use the notation $\spa{j}.{l}\equiv \langle
j^- | l^+ \rangle $, $\spb{j}.{l} \equiv \langle j^+ |l^- \rangle
$, with $| i^{\pm}\ra $ being  massless Weyl spinors with momentum
$k_i$ and chirality $\pm$~\cite{SpinorHelicity,ManganoReview}. The
spinor products are related to momentum invariants by
$\spa{i}.j\spb{j}.i=2k_i \cdot k_j\equiv s_{ij}$ with
$\spa{i}.j^*=\spb{j}.i$. As in twistor-space studies we define,
$$
\lambda_i \;=\; |  i^+\ra   
\; , \;\;\;
\bar\lambda_i \;= \; |  i^-\ra  
\,.
\equn
$$
Inspired by the duality between twistor string theory and
Yang-Mills~\cite{WittenTopologicalString} (and generalising a
previous description of the simplest gauge theory amplitudes by
Nair~\cite{Nair1988bq}), Cachazo, Svr\v{c}ek and Witten proposed a
reformulation of perturbation theory in terms of off-shell
MHV-vertices~\cite{Cachazo:2004kj}, which can be depicted,
\vspace{0.5cm}
\begin{center}
\begin{picture}(100,100)(55,-40)
\SetWidth{0.7} \Line(30,30)(0,30) \Line(30,30)(10,50)
\Line(30,30)(40,50) \Line(30,30)(10,10) \Line(30,30)(50,30)

\Line(120,30)(100,50) \Line(120,30)(140,50) \Line(120,30)(100,30)
\Line(120,30)(140,30) \Line(120,30)(120,10)

\Line(240,30)(220,30) \Line(240,30)(260,50) \Line(240,30)(260,10)
\Line(240,30)(270,30)


\Vertex(30,30){2} \Vertex(120,30){2} \Vertex(240,30){2}
\Text(-35,32)[c]{{\huge $\sum$}} \Text(7,0)[c]{$k_{i_1}^-$}
\Text(-9,32)[c]{$k_{i_2}^+$} \Text(7,61)[c]{$k_{i_3}^+$}
\Text(39,55)[c]{$\ldots$} \Text(58,29)[c]{$\times$}
\Text(93,29)[c]{$\times$}
\Text(75,29)[c]{$\displaystyle\frac{1}{p_{j_1}^2}$}
\Text(160,29)[c]{$\times\ \ \ldots$} \Text(212,29)[c]{$\times$}
\end{picture}
\end{center}
\vspace{-1.2cm}

The off-shell continuation for a leg of momentum $p$
was achieved by  replacing
$\lambda(p)$ by,
$$
\lambda_a(p )\;=\;  p_{a\dot a } \bar \eta^{\dot a}\,, \equn
$$
where $\bar \eta^{\dot a}$ is an arbitrary reference spinor.
While individual CSW diagrams depend on $\bar\eta$, the full amplitude is
$\bar\eta$-independent. This reformulation has led to or inspired a variety of
calculational advances, both for tree level scattering~\cite{trees}
 and at loop level~\cite{oneloop} in Yang-Mills theory.

This reformulation has been
demonstrated to reproduce all known results for gluon scattering at tree level
and, often,
gives relatively simple expressions for these amplitudes. Although originally
given for gluon scattering only, these rules have been shown to extend to
other types of massless particle~\cite{CSW:matter} and indeed to
massive particles~\cite{CSWmassive}.
It has been shown~\cite{Brandhuber:2004yw,Quigley:2004pw,Bedford:2004py}
that, with the correct off-shell prescription,
these vertices can be used to reproduce known one-loop results~\cite{BDDKa,BDDKb}
in supersymmetric theories.

In an alternate approach to computing tree level amplitudes, Britto,
Cachazo, Feng and Witten~\cite{Britto:new} obtained a recursion relation
based on analytically shifting a pair of external legs,
$$
\eqalign{
\lambda_i \;\longrightarrow\;  \lambda_i \;+\;z \lambda_j\,,
\;\;\;
\bar \lambda_j \;\longrightarrow\;  \bar\lambda_j \;-\;z  \bar\lambda_i\,,
\cr}
\equn
$$
and determining the physical amplitude, $A_n(0)$, from the poles in the
shifted amplitude, $A_n(z)$. This leads to a recursion relation in the number of external legs, $n$, of
the form,
$$
A_n(0) \;=\;  \sum_\alpha  \hat A_{n-k_\alpha+2}(z_\alpha)\times {
i \over P_\alpha^2}
         \times \hat A_{k_\alpha}(z_\alpha)\,,
\equn
$$
where the factorisation is only on these poles, $z_\alpha$,
 where legs $i$ and $j$ are connected to
different sub-amplitudes. This is depicted below:

\begin{center}
\begin{picture}(100,100)(15,-40)
\SetWidth{0.7}
\Line(10,30)(30,30)
\Line(30,50)(30,30)
\Line(45,45)(30,30)
\Line(15,45)(30,30)
\SetWidth{0.7}
\Line(55,30)(30,30)
\SetWidth{0.7}
\Vertex(26,40){2}
\SetWidth{1}
\Line(30,10)(30,30)
\SetWidth{0.7}
\GCirc(30,30){5}{0.8}
\SetWidth{0.7}
\Line(75,30)(100,30)
\SetWidth{0.7}
\Line(100,50)(100,30)
\Line(85,45)(100,30)
\Line(115,45)(100,30)
\Line(100,30)(120,30)
\Vertex(104,40){2}
\SetWidth{1}
\Line(100,10)(100,30)

\SetWidth{0.7}
\GCirc(100,30){5}{0.8}
\Text(65,29)[c]{$\displaystyle\frac{i}{P_\alpha^2}$}
\Text(31,0)[c]{{\rm $\hat k_i$}}
\Text(101,0)[c]{{\rm $\hat k_j$}}
\Text(-20,30)[c]{\Huge$\sum$}
\Text(-20,10)[c]{$\alpha$}
\end{picture}
\end{center}
\vspace{-0.8cm}

These recursion relations also give relatively compact formulae for
tree amplitudes~\cite{Luo:2005rx,SplitHelicity}.
Recursion relations based on analyticity can also be used at loop level both to
calculate rational terms~\cite{BDKrecursionA} and the coefficients of integral
functions~\cite{BBDI2005}.

The factorisation properties of the
amplitudes seem to lie at the heart of both approaches. In both
cases the amplitude is expressed as a sum of its factorisations in a
well specified manner. As such, one might hope to derive the
MHV-vertex formulation by applying an
analytic shift and obtaining a recursion relation.
In ref.~\cite{Kasper} it was demonstrated that such shifts exist and can be
used to derive the MHV vertex approach in gauge theory.
The shift
affects all of the negative helicity legs, $k_{m_i}$,
$$
\bar\lambda_{m_i}\; \to \;\hat{\bar\lambda}_{m_i}  \;=\;\bar\lambda_{m_i} \;+\;z r_i \bar \eta\,,
\equn\label{Kshift}
$$
with the $r_i$ chosen to ensure momentum conservation.

Most of the above developments have been made for gauge
theory amplitudes. 
The existence of a BCFW recusion relation for gravity amplitudes 
was strongly supported in~\cite{BBSTgravity,CSgravity}, 
and in this article we construct a 
CSW approach using the newly established shift (\ref{Kshift})
under the assumption that gravity amplitudes are sufficiently
well behaved for large values of $z$ in (\ref{Kshift}).

The key ingredient in obtaining the MHV rules is the analytic
structure of the amplitude which also underlies the derivation of the
recursion relations. In this context it becomes clear that these two
formalisms have their roots in the same physical behaviour of on-shell
amplitudes.

\section{Graviton Scattering Amplitudes}

Graviton scattering amplitudes are generally considerably
more complicated than those for gauge theory. To date,
explicit expressions have only been given for the MHV
amplitudes~\cite{BerGiKu,BBSTgravity} and for the six-point NMHV
amplitude~\cite{CSgravity,BDIgravity}. (As for gauge theories,
amplitudes with all helicities identical vanish, as do those with
one different, $M(1^{\pm},2^+,3^+,\cdots, n^+)=0$.)

In principle, gravity amplitudes can be constructed through the
Kawai, Lewellen and Tye (KLT)-relations~\cite{KLT}.
The explicit forms of these, up to six points, are,
$$
\eqalign{\hspace{1.2cm}
M_3^{\rm tree}(1,2,3) \;=\;
&
-iA_3^{\rm tree}(1,2,3)A_3^{\rm tree}(1,2,3)\,,
\cr
M_4^{\rm tree}(1,2,3,4) \;=\;
&
-is_{12}A_4^{\rm tree}(1,2,3,4)A_4^{\rm tree}(1,2,4,3)\,, \label{KLTFour} \cr
M_5^{\rm tree}(1,2,3,4,5) \;=\;&
\; is_{12}s_{34}\ A_5^{\rm tree}(1,2,3,4,5)A_5^{\rm tree}(2,1,4,3,5) \cr
& \hskip 1.9 cm \null
\;+\;i s_{13}s_{24}\ A_5^{\rm tree}(1,3,2,4,5)A_5^{\rm tree}(3,1,4,2,5)\,,
\label{KLTFive} \cr
M_6^{\rm tree}(1,2,3,4,5,6) \;=\; &
-is_{12}s_{45}\ A_6^{\rm tree}(1,2,3,4,5,6)(s_{35}A_6^{\rm tree}(2,1,5,3,4,6)
\cr
& \hskip 1.9 cm \null
\;+\;(s_{34}+s_{35})\ A_6^{\rm tree}(2,1,5,4,3,6)) \;+\; {\cal P}(2,3,4)\,,
\cr}
\equn\label{KLTSix}
$$
where
${\cal P}(2,3,4)$ represents the sum over
permutations of legs $2,\ 3,\ 4$ and the $A^\tree_n$ are the tree-level
colour-ordered gauge theory partial amplitudes. We have suppressed
factors of $g^{n-2}$ in the $A^\tree_n$, as well as a factor
of $(\kappa/2)^{n-2}$ in the gravity amplitude.

This formulation allows results from Yang-Mills theory to be recycled
in theories of gravity and
supergravity~\cite{StringBased,Bern:1998xc,BDWGravity,GravityReview,EffKLT}.
While these relations are directly applicable to tree amplitudes, this
formulation also has implications for loop amplitude calculations,
particularly in unitarity based methods where the tree amplitudes are
used to compute the loop amplitudes~\cite{BDDPR,Bern:1998sv,DunNor}.
Consequently, similar relationships can hold for the coefficients of
integral functions~\cite{BeBbDu}.

Although, in principle, the KLT relations can be used to calculate gravity
tree amplitudes, they have several undesirable features. Firstly, the
factorisation structure is rather obtuse.  The Yang-Mills tree
amplitudes contain single poles so we might expect un-physical double
poles to appear in the sum. These are actually canceled by the
multiplying momentum factors, but often in a non-trivial manner.
Secondly, the expressions do not tend to be compact as the permutation sums
grow rather quickly with the  number of points. In fact,
the Berends, Giele and Kuijf (BGK)
form of the MHV gravity amplitude~\cite{BerGiKu},
$$
\eqalign{
M_n^{\rm tree}
&(1^-,2^-,3^+,\cdots, n^+)
\cr
&\;=\;-i\spa1.2^8\times
\biggl[{ \spb1.2\spb{n-2}.{n-1}    \over \spa1.{n-1} N(n) }
\Bigl(  \prod_{i=1}^{n-3} \prod_{j=i+2}^{n-1} \spa{i}.j \Bigr)
\prod_{l=3}^{n-3} (-[n|K_{l+1,n-1}|l\ra)
\cr&\hspace{8.7cm}+{\cal P}(2,3,\cdots,n-2)
\biggr]\,,
\cr}
\equn\label{BGKform}
$$
is rather more compact than that of the KLT sum (as is the expression
in~\cite{BBSTgravity}.)
In the above we use the definitions,
$$
\BBRQ k {{K}_{i,j}} l \;\equiv\; \BRQ {k^+} {\Slash{K}_{i,j}} {l^+} \;\equiv\; \BRQ
{l^-} {\Slash{K}_{i,j}} {k^-} \;\equiv\; \la l |{{K}_{i,j}}| k] \;\equiv\; \sum_{a=i}^j\spb k.a\spa a.l\,, \equn
$$
and $N(n)=\prod_{1\leq i<j \leq n} \spa{i}.{j}$.

Both the KLT form of the MHV amplitude~(\ref{KLTSix}) and the above
form~(\ref{BGKform}) display a feature not shared by the Yang-Mills
expressions: they not only depend on the holomorphic variables
$\lambda$, but also on the $\bar\lambda$ - within the $s_{ij}$ for the
KLT expression and explicitly in the BGK expression. In both cases
this dependence is polynomial in the numerator. This feature
complicates the twistor space structure of any potential form of a MHV
vertex for gravity.  For Yang-Mills, the holomorphic vertex
corresponds simply to points lying on a line in twistor space.  For
gravity the picture will be of points lying on the ``derivative of a
$\delta$-function''~\cite{WittenTopologicalString}. The practical
difference is that both $\lambda(q)$ and $\bar\lambda(q)$ must be
correctly continued off-shell.  (The exception to this is the
three-point vertex for which the gravity MHV expression is
holomorphic.)  Various attempts have been made to find the off-shell
continuation~\cite{GBgravity,ZhuGrav}.  Despite the failure to find a
MHV vertex formulation, gravity amplitudes are amenable to recursive
techniques~\cite{BBSTgravity,CSgravity}.  In~\cite{NairGravity} a
current algebra formulation was demonstrated for the MHV gravity
amplitudes which also suggests that a MHV vertex might exist.

\section{NMHV Graviton Scattering Amplitudes}
We shall demonstrate the off-shell MHV vertex for gravity using the
analytic structure of the amplitudes with three negative helicity legs
(known as ``next-to-MHV`` or NMHV amplitudes).  The shift
of~\cite{Kasper} allows us to rewrite the NMHV amplitudes as products
of MHV-amplitudes and thus gives a CSW type expansion for these
amplitudes directly, from which we can identify the off-shell gravity
MHV-vertices.

Let us start by considering a generic $n$-point NMHV graviton amplitude
$M_n(m_1^-,\break m_2^-, m_3^-, \cdots ,n^+)$, where
we label the three negative helicity legs $1$, $2$ and $3$ by $m_i$.
We can make the same continuation as in the Yang-Mills case,
$$
\eqalign{
\hat{\bar\lambda}_{m_1}  \;=\;\bar\lambda_{m_1} \;+\;z \spa{m_2}.{m_3} \bar
\eta\,,
\cr
\hat{\bar\lambda}_{m_2}  \;=\;\bar\lambda_{m_2} \;+\;z \spa{m_3}.{m_1} \bar
\eta\,,
\cr
\hat{\bar\lambda}_{m_3}  \;=\;\bar\lambda_{m_3} \;+\;z \spa{m_1}.{m_2} \bar
\eta\, ,
\cr}
\equn\label{ourshift}
$$
which shifts the momentum of the negative helicity legs,
$$
\hat{k}_{m_i}(z) \;=\;  \lambda_{m_i}\left(\bar \lambda_{m_i}
\;+\;z\spa{m_{i-1}}.{m_{i+1}} \bar \eta\right)\,,
\equn
$$
but leaves them on-shell, $k_{m_i}(z)^2=0$, while the combination
$k_{m_1}(z)+k_{m_2}(z)+k_{m_3}(z)$ is independent of $z$.
Under the shift we obtain the analytic continuation of the amplitude
$M_n(z)=\hat M_n$
into the complex plane. We use a `hat` to distinguish the
unshifted objects, $a$, from the shifted ones, $\hat a$.

For a shifted amplitude we can evaluate the following
contour integral at infinity,
$$
\frac{1}{2\pi i} \oint {dz\over z}M_n(z) \;=\; C_\infty
\;=\; M_n(0) + \sum_\alpha  {\rm Res}_{z=z_\alpha}{M_n(z) \over z}\,.
\equn
$$
If $M_n(z)$ is rational with simple poles at points $z_\alpha$
and $C_\infty$ vanishes,
$M_n(0)$
can be expressed in terms of residues,
$$
M_n(0)  \;=\; -\sum_{\alpha} \, {\rm Res}_{z=z_\alpha} {M_n(z)\over z }\,.
\equn
$$
The first condition is satisfied as a result of the general
factorisation properties of amplitudes, however the second is
difficult to prove in general
for gravity amplitudes.

The shifted amplitude has poles in $z$ whenever a momentum invariant
$\hat P^2(z)$ vanishes. Given the form of the shift, all momentum
invariants apart from those containing all or none of the negative
helicities are $z$-dependent.  Thus the NMHV amplitudes have
factorisations where two of the negative helicity legs lie on one
side and one on the other.  For the above shift it can be checked that
all factorisations involving the MHV googly 3-point amplitude $M(-++)$
vanish. All poles of the amplitude must therefore factorise as,
$$
M^{\rm MHV}(m_{i_1}^-,\cdots , P^-) \times { i \over P^2 }\times M^{\rm
MHV}((-P)^+, m_{i_2}^-,m_{i_3}^-,\cdots )\,,
\equn
$$
for $i_k=1,\ 2$ or $3$, as we expect for a CSW-type expansion.
$\hat P^2(z)$ vanishes linearly in $z$, so  $M_n(z)$ has simple
poles when $z_\alpha$ satisfies,
$$
\hat P^2  \;=\; P^2 +z_\alpha\spa{m_{i_2}}.{m_{i_3}} [\eta | P | m_{i_1}
\ra=0\,.
\equn
$$

The residue at each pole is just the product of the two MHV tree
amplitudes  evaluated at $z=z_\alpha$.
Spinor products $\spa{i}.{j}$ which are not $z_\alpha$ dependent take
their normal values, while
terms like $\la{i}{\hat P}\ra$ are evaluated by noting,
$$
\la{i}\,{\hat P}\ra \;=\;
{\la i\,{\hat P}\ra[{\hat P}\,{\eta}] \over [{\hat P}\,{\eta}]  }
\;=\;  { \la i | P |  \eta ] \over \omega}\,,
\equn\label{NMHV2}
$$ where $P$ is the unshifted form.
The objects $\omega$ will cancel between the two tree amplitudes since the
product has zero spinor weight in $P$.  This substitution is precisely the
CSW prescription,
$\lambda(P) \longrightarrow P|\eta]$.

For Yang-Mills this would be all we need, but for gravity we must also
consider
substitutions for $\spb{i}.j$ where $i$ and/or $j$ are one of the negative
helicities or $\hat P$. These substitutions are
of the form,
$$
\eqalign{
[{l^+}\,{\hat P}]
\;=\;& {[l^+\,{\hat P}] \la{\hat P}\,\alpha\ra
\over \la{\hat P}\,\alpha\ra}
\;=\; { \omega  [ l^+ | \hat P | \alpha \ra \over [ \eta | P | \alpha \ra }
\;=\; { \omega   [ l^+ | P | m_{i_1} \ra
   \over [ \eta | P | m_{i_1} \ra }\,,
\cr
\spb{\hat m_{i_2}}.{\hat m_{i_3}}
\;=\;& \spb{m_{i_2}}.{m_{i_3}}  + z_\alpha [ \eta | P_{m_{i_2}m_{i_3}} |
m_{i_1} \ra\,,
\cr
\spb{\hat m_{i_1}}.{l^+}\,
\;=\;&  \spb{m_{i_1}}.{l^+} +z_\alpha \spb{\eta}.{l^+}
\spa{m_{i_2}}.{m_{i_3}}\,,
\cr}
\equn\label{NMHV1}
$$
where $l^+$ denotes a positive helicity leg. We choose the arbitrary spinor
$\alpha$
to be $m_{i_1}$ in order to replace $\hat P$  by $P$.
Equations (\ref{NMHV2}) and (\ref{NMHV1}) are the specific
substitutions that determine the value
of the MHV amplitudes on the pole and thus the MHV vertices.

Note that the form of the off-shell continuation,
$$
\hat{\bar\lambda}_{m_1} = \bar\lambda_{m_1}+ z\spa{m_2}.{m_3} \bar \eta
 = \bar\lambda_{m_1}- { P^2 \bar \eta
\over  [ \eta | P | {m_1} \ra}\, ,
\equn
$$
can be interpreted as yielding contact terms
since the  $P^2$ factor may  cancel the pole.

We conclude that the NMHV graviton scattering amplitude can be expressed
in terms of MHV vertices as,
$$\hspace{-0.3cm}
\eqalign{
&M_n(1^-,2^-,3^-,4^+,\ldots, n^+)
\;=\;
\sum_{r=0}^{n-4}
\sum_{{\cal P}(i_1,i_2,i_3) }\sum_{{\cal P}(d_i) }
M_{r+3}^{{\rm MHV}}(\hat m_{i_2}^-,\hat
m_{i_3}^-,d_1^+,\cdots, d_r^+,\hat P^+ )(z_r)
\cr&\hspace{5cm}\times { i \over P_{m_{i_1} d_{r+1}\cdots, n}^2 }
\times
M_{n-r-1}^{{\rm MHV}}((-\hat P)^-,\hat m_{i_1}^-,d_{r+1}^+,\cdots,n
)(z_r)\,.
\cr}
\equn
$$
Here the sums over ${\cal P}_r(i_1,i_2,i_3)$ and ${\cal P}_r(d_i)$ are respectively
sums over those permutations of the negative and positive helicity legs that swap legs between the
two MHV vertices.

We now turn to the discussion of the behaviour of $M_n(z)$ for large $z$.
By naive power counting one might expect shifted gravity amplitudes to
diverge at large $z$.  However in both ref.~\cite{BBSTgravity}
and ref.~\cite{CSgravity} it was established by various techniques,
including numerical studies, that NMHV gravity amplitudes do vanish asymptotically
under the BCFW shift. This behaviour is difficult to prove either by
analysing Feynman diagrams or using the KLT relations, since large
cancellations are inherent in both formalisms.
Under the shift~(\ref{ourshift}) the amplitudes we have examined
are very well behaved at large $z$, with
$$
M_{6,7}(z) \;\sim\; {1 \over z^5}\,,
\equn
$$
for both the six and seven point NMHV amplitudes. This is a much stronger
behaviour than under the BCF shift where,
$
M_{6,7}(z)\;\sim\; {1 \over z}\,.
$

If we choose a specific value for the reference spinor,
$$
\bar\eta = \bar \lambda_a\,,
\equn
$$
where
$a$ is one of the positive helicity legs, then the shift we
use is a combination of three BCF shifts involving the three negative
helicity legs and a positive helicity leg $a$,
$$
\eqalign{
\lambda_a \longrightarrow \lambda_a +z_1 \spa{2}.3 \lambda_{1}
,\;\;\;
\bar \lambda_1 \longrightarrow \bar \lambda_1-z_1\spa{2}.3\bar\lambda_a \, ,
\cr
\lambda_a \longrightarrow \lambda_a +z_2 \spa{3}.1 \lambda_{2}
,\;\;\;
\bar \lambda_2 \longrightarrow \bar \lambda_2-z_2\spa{3}.1\bar\lambda_a \, ,
\cr
\lambda_a \longrightarrow \lambda_a +z_3 \spa{1}.2 \lambda_{3}
,\;\;\;
\bar \lambda_3 \longrightarrow \bar \lambda_3-z_3\spa{1}.2\bar\lambda_a \, ,
\cr}
\equn
$$
with $z_1=z_2=z_3$. The shift on $\lambda_a$ vanishes due to the
Schouten identity.  In ref.~\cite{CSgravity} it was proven that the
amplitude vanishes at infinity under a single shift of this form,
providing further evidence that the NMHV amplitude vanishes
asymptotically under the shift~(\ref{ourshift}).

\subsection{Five-Point Example $M(1^-,2^-,3^-,4^+,5^+)$}
In this section we show, using an explicit example, how MHV vertices
can be assembled into graviton scattering amplitudes.  The first
non-trivial example is the five-point amplitude,
$M(1^-,2^-,3^-,4^+,5^+)$.  This is a 'googly' amplitude as it can be
obtained by conjugating the five-point MHV amplitude.  As above we
shift the negative helicity legs, here $k_1,k_2$ and $k_3$, and compute
the residues of the amplitude $\, M(\hat 1^-,\hat 2^-,\hat
3^-,4^+,5^+)(z)$.  The expansion in terms of MHV vertices is
non-trivial and reveals the structure of the MHV vertices.

Up to relabeling we have two types of residue,
$$
\hspace{1.5cm}
\eqalign{ D_1(1^-,2^-,3^-,4^+,5^+)\;=\; & M(\hat 2^-,\hat 3^-,\hat
p^+)\times { i\over s_{23} } \times M((-\hat p)^-,4^+,5^+,\hat 1^-)\,,
\cr D_2(1^-,2^-,3^-,4^+,5^+)\;=\; &
M(\hat 2^-,\hat 3^-,4^+,\hat p^-)\times { i\over
s_{15} } \times M((-\hat p)^+,5^+,\hat1^-)\,, \cr} \equn
$$
which can be associated to the CSW diagrams,
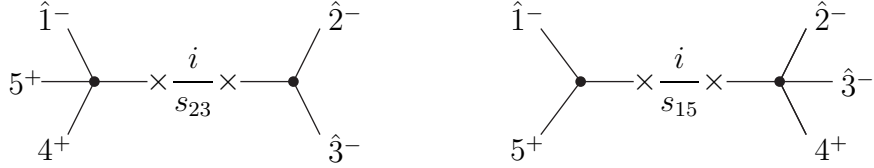
\begin{center}
\begin{picture}(100,100)(55,-40)
\Vertex(30,30){2}
\Vertex(105,30){2}
\Line(30,30)(50,30)
\Line(85,30)(105,30)
\Line(30,30)(10,30)
\Line(30,30)(20,50)
\Line(30,30)(20,10)
\Line(105,30)(115,50)
\Line(105,30)(115,10)
\Text(125,55)[c]{$\hat 2^-$}
\Text(125,5)[c]{$\hat 3^-$}
\Text(4,30)[c]{$5^+$}
\Text(15,55)[c]{$\hat 1^-$}
\Text(15,5)[c]{$4^+$}
\Text(67.5,30)[c]{$\displaystyle\times\frac{i}{s_{23}}\times$}
\end{picture}
\begin{picture}(100,100)(-25,-40)
\Vertex(30,30){2}
\Vertex(105,30){2}
\Line(105,30)(85,30)
\Line(50,30)(30,30)
\Line(105,30)(125,30)
\Line(105,30)(115,50)
\Line(105,30)(115,10)
\Line(105,30)(115,50)
\Line(105,30)(115,10)
\Line(30,30)(15,10)
\Line(30,30)(15,50)
\Text(10,55)[c]{$\hat 1^-$}
\Text(10,5)[c]{$5^+$}
\Text(135,30)[c]{$\hat 3^-$}
\Text(125,55)[c]{$\hat 2^-$}
\Text(125,5)[c]{$4^+$}
\Text(67.5,30)[c]{$\displaystyle\times\frac{i}{s_{15}}\times$}
\end{picture}
\end{center}
\vspace{-1.2cm}
Explicitly we find for the three-point function,
$$
M(\hat 2^-,\hat 3^-,\hat p^+) \;=\; i
{\spa{2}.{3}^6 \over \spa{2}.{\hat p}^2\spa{\hat p}.{3}^2 }
= i {\omega^4 \spa{2}.{3}^6 \over [\eta | P_{23} | 2 \ra^2
[\eta | P_{23} | 3 \ra^2 }\,.
\equn
$$
The four point amplitude can be expressed in several ways, including,
$$
\eqalign{ M_4((-\hat p)^-,4^+,5^+,\hat 1^-) &\;=\;  -is_{45}
A_4((-\hat p)^-,4^+,5^+,\hat 1^-)A((-\hat p)^-,5^+,4^+,\hat 1^-)
\cr &={ i\,s_{45} \spa{ \hat p}.1^8 \over \spa{\hat p}.{4}
\spa{4}.5\spa{5}.1 \spa{1}.{\hat p} \spa{\hat p}.{5}
\spa{5}.4\spa{4}.1 \spa{1}.{\hat p}
   }
\cr
&\;=\;  { i\, \spb4.5 \over \spa{1}.4\spa{1}.5 \spa{4}.5 }
{\omega^{-4}
 [\eta | P_{23} | 1 \ra^6  \over
[\eta | P_{23} | 4 \ra [\eta | P_{23} | 5 \ra}\,,
\cr}
\equn
$$
giving the tree diagram as,
$$
D_1(1^-,2^-,3^-,4^+,5^+)={ i\, \spb4.5
 [\eta | P_{23} | 1 \ra^6  \over\spa{1}.4\spa{1}.5 \spa{4}.5
[\eta | P_{23} | 4 \ra [\eta | P_{23} | 5 \ra} \,{i\over s_{23}}
\, { i\spa{2}.{3}^6 \over [\eta | P_{23} | 2 \ra^2 [\eta | P_{23}
| 3 \ra^2 }\,. \equn
$$
For this particular diagram the prescription implied by the shift
 is equivalent to the CSW rules for gauge theory as there is no need to find a
continuation for $\bar\lambda$.

For $D_2$ we find,
$$
D_2(1^-,2^-,3^-,4^+,5^+)\;=\; {  i\,\spa{2}.{3}^7 \spb{\hat 2}.{\hat
3} \over \spa{2}.4 \spa{3}.4 [\eta | P_{15} | 4 \ra^2 [\eta |
P_{15} | 2 \ra [\eta | P_{15} | 3 \ra
 }
\,{i\over s_{15}}\, {i\,[\eta | P_{15} |5\ra^6 \over [\eta |
P_{15} | 1 \ra^2 \spa{5}.1^2 }\,. \equn
$$
This differs from the simple CSW prescription in the
definition of $\spb{\hat 2}.{\hat 3}$.
Here,
$$
\eqalign{ \spb{\hat 2 }.{\hat 3} \;=\; &\spb{2}.{3} +z \left(
\spa{3}.{1} \spb{\eta}.3 +\spa{1}.{2} \spb{2}.{\eta} \right)
\;=\;\spb{2}.{3} + z [\eta | P_{23} |1 \ra \cr
&\;=\;\spb{2}.{3} -{
P_{15}^2  \over \spa{2}.3 [\eta | P_{15} | 1 \ra  }
 [\eta | P_{23} |1 \ra\,.
\cr}
\equn
$$
With this substitution we can verify that the sum of diagrams is
independent of $\bar\eta$ and equal to the conjugate
of the five-point MHV tree amplitude.

For the six-point amplitude there are three diagrams. We have
explicitly checked that the sum over permutations of these
diagrams is equal to the known form of the six-point NMHV
amplitude~\cite{CSgravity,BDIgravity}. Seven point
NMHV amplitudes can be obtained explicitly using the
KLT relationships - at least using computer algebra. We have
checked numerically that the seven-point amplitudes obtained from
the MHV vertices match those obtained from the KLT relation.

\subsection{Remarks on the Twistor Space Structure of MHV Gravity Amplitudes}
The twistor space structure of an amplitude refers to the support of the amplitude
after it has been Penrose transformed into twistor space variables.
As was shown in ref.~\cite{WittenTopologicalString}, the twistor space
support of an amplitude can be tested by simply acting with certain differential operators,
without having to resort to Penrose or Fourier transformations. The operator of particular
interest is the 'collinearity operator',
$$
[F_{ijk} , \eta ] \;=\;
 \spa{i}.j \left[{ \partial\over \partial\bar\lambda_k},\eta\right]
\;+\;\spa{j}.k \left[{ \partial\over
\partial\bar\lambda_i},\eta\right] \;+\;\spa{k}.i \left[{
\partial\over
\partial\bar\lambda_j},\eta\right]\,.
\equn
$$

The expressions for the NMHV gravity amplitudes can be used to test the
twistor structure of gravity amplitudes. In this case,
MHV amplitudes are annihilated by multiple applications of the collinearity operator,
$$
F^h M_n^{\rm MHV}\;=\;0\,,
\equn
$$
for some $h$.  This is interpreted as the support being non-zero only if the
points are ``infinitesimally'' close to a line in twistor
space~\cite{WittenTopologicalString}.
In ref.~\cite{BeBbDu} it was explicitly shown that,
$$
[F_{ijk},\eta]^{n-2} M_n^{\rm MHV} \;=\;0\,,
\equn
\label{lotsofF}
$$
for n-point amplitudes with $n \leq 8$. If we compare the action of the collinearity operator
on the amplitude with that of the shift,
$$
\eqalign{
\bar\lambda_{i}  \;\to\;\bar\lambda_{i} \;+\;z \spa{j}.{k} \bar \eta\,,
\cr
\bar\lambda_{j}  \;\to\;\bar\lambda_{j} \;+\;z \spa{k}.{i} \bar \eta\,,
\cr
\bar\lambda_{k}  \;\to\;\bar\lambda_{k} \;+\;z \spa{i}.{j} \bar \eta\,,
\cr}
\equn
$$
it can be seen that,
$$
[F_{ijk},\eta] M_n(0) \;=\; \frac{\partial}{\partial z}\hat M_n(z)|_{z=0}\,.
\equn
$$
Equation (\ref{lotsofF}) can thus be understood in terms of the
number of $s_{ij}$ factors in the KLT form of the amplitude: each
factor of $s_{ij}$ can introduce at most one power of $z$ and
in the KLT form there are $n-3$ factors of $s_{ij}$ in the n-point amplitude, so
$n-2$ applications of $[F_{ijk},\eta]$ are sufficient to annihilate the
amplitude.

\section{Beyond NMHV}
In this section we will illustrate a generalisation of the
CSW-rules for gravity amplitudes with an example and verify the
rules for this amplitude. Finally, we will present the generic
rules and discuss their proof via the BCFW approach.

\subsection{General CSW rules}
We will now extend the CSW rules for NMHV amplitudes into
more generic rules for the expansion of N$^n$MHV amplitudes,
that is amplitudes with $n+2$ negative helicity legs and the rest positive.

Consider the N$^n$MHV amplitude with $N$ external legs. One would, as
in the Yang-Mills case, begin by drawing all diagrams which may be
constructed using MHV vertices. For the off-shell continuation, three-point
MHV vertices are non-vanishing.  The contribution from each diagram
will be a product of $(n+1)$ MHV vertices and $n$
propagators as indicated below.
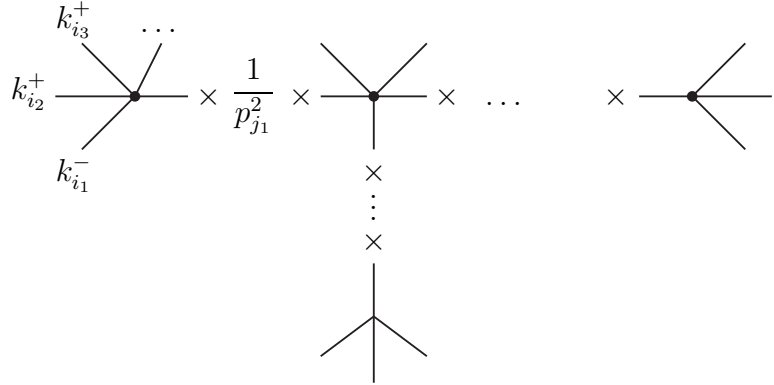
\begin{center}
\begin{picture}(100,100)(55,-40)
\SetWidth{0.7}
\Line(30,30)(0,30)
\Line(30,30)(10,50)
\Line(30,30)(40,50)
\Line(30,30)(10,10)
\Line(30,30)(50,30)

\Line(120,30)(100,50)
\Line(120,30)(140,50)
\Line(120,30)(100,30)
\Line(120,30)(140,30)
\Line(120,30)(120,10)

\Line(240,30)(220,30)
\Line(240,30)(260,50)
\Line(240,30)(260,10)
\Line(240,30)(270,30)

\Line(120,-53)(120,-78)
\Line(120,-53)(100,-68)
\Line(120,-53)(140,-68)
\Line(120,-53)(120,-33)

\Vertex(30,30){2}
\Vertex(120,30){2}
\Vertex(240,30){2}
\Text(7,1)[c]{$k_{i_1}^-$}
\Text(120.5,1)[c]{$\times$}
\Text(120.5,-9)[c]{$\vdots$}
\Text(120.5,-25)[c]{$\times$}
\Text(-10,32)[c]{$k_{i_2}^+$}
\Text(7,59)[c]{$k_{i_3}^+$}
\Text(39,55)[c]{$\ldots$}
\Text(58,30)[c]{$\times$}
\Text(93,30)[c]{$\times$}
\Text(75,30)[c]{$\displaystyle\frac{1}{p_{j_1}^2}$}
\Text(160,30)[c]{$\times\ \ \ldots$}
\Text(212,30)[c]{$\times$}
\end{picture}
\end{center}
\vspace{1.2cm}
In contrast to gauge theory the CSW diagrams for gravity have no cyclic ordering of
the external legs.

We denote internal momenta by $p_j$ for $j=1,...,n$ and external momenta by $k_i$ for $i=1,...,N$.
We label the vertices by $l$ for $l=1,...,(n+1)$. The momenta leaving
MHV vertex $l$ are collected into the set $K_l$ and the number of external legs
of MHV vertex $l$ will be denoted by $N_l$.

The contribution of a given diagram to the total amplitude can be calculated by
evaluating the product of MHV amplitudes and propagators,
$$
M^n_N{\big |}_{\mbox{\small CSW-diagram}}\;=\;\left(\prod_{l=1,n+1} M^{{\rm MHV}}_{N_l}(\hat K_l)\right)
\prod_{j=1,n} \frac{i}{ p_{j}^2}\, ,
\equn\label{CSWrule}
$$
where the propagators are computed on the set of momenta $k_i$ and $p_j$, and the MHV vertices are evaluated
for the momenta $\hat k_i$ and $\hat p_j$. The definitions of these momenta are given below.

The momenta $k_i$ are the given external momenta and the internal momenta, $p_{j}$, are given by
momentum conservation on each MHV-vertex.

The momenta $\hat k_i$ and $\hat p_{j}$ are uniquely specified so that they are massless
and  obey momentum conservation constraints at each vertex.
Explicitly they are given by shifting the negative helicity
legs $k_{i^-}$,
$$
\hat{k}_{i^-} \;=\; k_{i^-} \;+\; a_{i^-} \lambda_{(i^-)} \bar\eta \,,
\equn
$$
and leaving the positive helicity legs $k_{i^+}$ untouched. This introduces
$n+2$ parameters, $a_{i^-}$.
Overall momentum conservation is used to fix two of these
parameters. Momentum conservation at each vertex
then gives the  momenta $\hat p_{j}$ as functions of $k_i$ and $a_i$.
Finally the remaining parameters are fixed such that all
internal momenta, $\hat p_{j}$, are massless,
$$
\hat p_{j}^2=0\,.
\equn
$$
This gives $n$ further linear constraints which are
sufficient to fix  the remaining $a_i$ uniquely
 for a given spinor $|\eta]$.
The MHV vertices in (\ref{CSWrule}) can then be
evaluated on the on-shell momenta $\hat k$ and $\hat p$.

\subsection{Example}
As an example we will discuss an explicit CSW diagram that contributes to the 8-point amplitude
$M(1^-,2^-,3^+,4^+,6^+,7^-,8^-)$.  The diagram is given by,
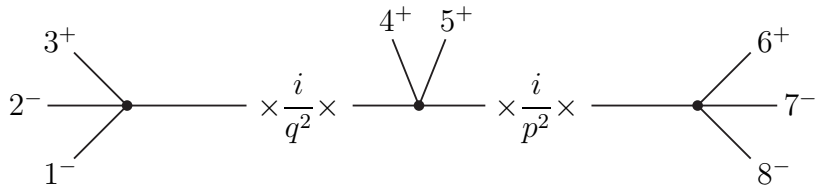
\begin{center}
\begin{picture}(100,100)(75,-30)
\SetWidth{0.7}
\Vertex(30,30){2}
\Vertex(140,30){2}
\Vertex(245,30){2}
\Line(30,30)(75,30)
\Line(115,30)(165,30)
\Line(205,30)(245,30)
\Text(95,29)[c]{$\times\displaystyle\frac{i}{q^2}\times$}
\Text(185,29)[c]{$\times\displaystyle\frac{i}{p^2}\times$}
\Line(140,30)(130,55)
\Line(140,30)(150,55)
\Line(30,30)(0,30)
\Line(30,30)(10,50)
\Line(30,30)(10,10)
\Line(245,30)(275,30)
\Line(245,30)(265,50)
\Line(245,30)(265,10)
\Text(5,5)[c]{$1^-$}
\Text(-8,31)[c]{$2^-$}
\Text(5,55)[c]{$3^+$}
\Text(132,63)[c]{$4^+$}
\Text(155,63)[c]{$5^+$}
\Text(275,55)[c]{$6^+$}
\Text(285,31)[c]{$7^-$}
\Text(275,5)[c]{$8^-$}
\end{picture}
\end{center}
\vspace{-1.1cm}
This specific diagram is interesting  as none of its
vertices can be written purely in terms of angles, $\spa{\,}.{\,}$.

Following the algorithm from above the diagram
contributes,
$$
\eqalign{
& M(\hat 1^-,\hat 2^-,3^+,\hat q^+)\frac{i}{q^2}M((-\hat
q)^-,4^+,5^+,(-\hat p)^-)\frac{i}{p^2}M(\hat p^+,6^+,\hat
7^-,\hat 8^-)\;=\;
\cr
&\hspace{-0.4cm}\frac{i\,\spa{\hat 1}.{\hat 2}^6\spb 3.{\hat q}}{\spa{\hat
2}.3\spa3.{\hat q}\spa {\hat q}.{\hat 1}\spa{\hat 2}.{\hat q}\spa
3.{\hat 1}}\,\frac{i}{q^2}\,\frac{i\,\spa {\hat p}.{\hat q}^6\spb
4.5}{\spa {\hat q}.4\spa4.5\spa5.{\hat p}\spa {\hat
q}.5\spa4.{\hat p}}\frac{i}{p^2}\frac{i\,\spa{\hat 7}.{\hat
8}^6\spb {\hat p}.6}{\spa{\hat 8}.{\hat p}\spa {\hat
p}.6\spa6.{\hat 7}\spa{\hat 8}.6\spa {\hat p}.{\hat
7}}\,.
\cr}
\equn
\label{example}
$$
The internal momenta, $q$ and $p$, are given by momentum
conservation: $q+k_1+k_2+k_3=0$ and $p+k_6+k_7+k_8=0$.

For the momenta $\hat k_i,\,i=1,2,7,8$, the shift is,
$$
\eqalign{
&\hat k_i\;=\;k_i\;+\;a_i\,\lambda_i\bar\eta,\quad i\;=\;1,2,7,8\,,
\cr}
\equn
$$
while the momenta $k_i$ with $i\;=\;3,4,5,6$ are untouched.
The momenta $\hat p$ and $\hat q$ and the parameters $a_i$ have to be
fixed such that the momentum flowing through each of the vertices is
preserved,
$$
\eqalign{
\sum_{i=1,8}\hat k_i\;=\;0\,,&\cr
\hat q\;+\;\hat k_1\;+\;\hat k_2\;+\;k_3\;=\;0\,,&\cr
\hat p\;+\;k_6\;+\;\hat k_7\;+\;\hat k_8\;=\;0\,.&
\cr}
\equn
$$
This leaves two free parameters which are
fixed such that,
$$
\hat q^2\;=\;0,\quad \hat p^2\;=\;0\,.
\equn
$$
There is a specific shift for each  CSW diagram. In
general, different diagrams that contribute to the same amplitude
yield different values of $a_i$.

The various spinor products in (\ref{example}) can be computed
using the above conditions for the $a_i$  and one finds expressions
very reminiscent of those derived by CSW for gauge theory,
$$
\eqalign{
&\spa{k_i}.{\hat q}\;=\;\spab k_i.(-P_{123}).\eta/\omega_q\,,\quad\omega_q\;=\;\spb{\hat q}.\eta\,,\cr
&\spa{k_i}.{\hat p}\;=\;\spab k_i.(-P_{678}).\eta/\omega_p\,,\quad\omega_p\;=\;\spb{\hat p}.\eta\,,\cr
&\spb3.{\hat q}\;=\;-\frac{\omega_q \spba3.{\hat P_{123}}.{\hat
p}}{\spba\eta.{\hat P_{123}}.{\hat p}}
\;=\;-\frac{\omega_q \spbb3.{\hat P_{123}\hat P_{678}}.\eta}{\spbb\eta.{\hat P_{123}\hat P_{678}}.\eta}\;=\;
\frac{\omega_q \spbb3.{P_{45}P_{678}}.\eta}{\spbb\eta.P_{123}P_{678}.\eta}\,,\cr
&\spb6.{\hat p}\;=\;-\frac{\omega_p \spba6.{\hat P_{678}}.{\hat
q}}{\spba\eta.{\hat P_{678}}.{\hat q}} \;=\;-\frac{\omega_p
\spba6.{\hat P_{678}\hat P_{123}}.\eta}{\spba\eta.{\hat P_{678}\hat P_{123}}.\eta} \;=\;
\frac{\omega_p
\spbb6.{P_{45}P_{123}}.\eta}{\spbb\eta.P_{678}P_{123}.\eta}\,.
\cr}
\equn
$$
Overall the $\omega_q$ and $\omega_p$ factors cancel  and we find that (\ref{example}) is given by,
$$
\eqalign{
&\hspace{-0.4cm}\frac{i\,\spa1.2^6\frac{\spbb 3.{
P_{45}P_{678}}.\eta}{\spbb\eta.P_{123}P_{678}.\eta}}{\spa2.3\spab3.P_{123}.\eta\spba\eta.P_{123}.1\spab2.P_{123}.\eta\spa3.1}
\,\frac{i}{t_{123}}\,\frac{i\,\spbb\eta. P_{678}P_{123}.\eta
^6\spb4.5}{\spba
\eta.P_{123}.4\spa4.5\spab5.P_{678}.\eta\spba\eta.P_{123}.5\spab4.P_{678}.\eta}\,
\cr
&\times\frac{i}{t_{678}}\,\frac{i\,\spa7.8^6\frac{\spbb\eta.P_{123}
P_{45}.6}{\spbb\eta.P_{123}P_{678}.\eta}}{\spab8.P_{678}.\eta\spba\eta.P_{678}.6\spa6.7\spa8.6\spba\eta.P_{678}.7}\,.
\cr}
\equn
$$
The rules used to compute the above N$^2$MHV diagram are a natural
generalisation of the NMHV-case. As we will discuss below, they
follow from BCFW recursions, providing the shifted amplitudes
vanish for large $z$.

\subsection{Proof of MHV-vertex rules}

In this section we prove the validity of the CSW-like  expansion
of the graviton scattering amplitudes in terms of MHV vertices with
the substitution rules of the previous section.  We will employ
a recursive proof analogous to that for Yang-Mills~\cite{Kasper} where
recursion was employed upon the number of minus legs in the tree
amplitudes.  As a first step we will prove the N${}^2$MHV case where
four legs have negative helicity and later we will generalise the proof for
generic N$^n$MHV amplitudes.

\subsection{MHV-vertex Expansion for  N${}^2$MHV Amplitudes}

We shall derive the CSW-like expansion for this amplitude by
factorising the amplitude in two steps. First we shall factorise the
amplitude into a product of MHV and NMHV amplitudes and then
factorise the NMHV amplitudes to complete the expansion.

We first apply a holomorphic shift
similar to the one discussed in \cite{Kasper} for gauge theory,
$$
\bar\lambda_i \longrightarrow
\bar\lambda_i+z_1 r_i^{(1)} \bar\eta
\; , \;\; i=1,\cdots,4\,,
\equn
$$
where the $r_i^{(1)}$ are restricted by momentum conservation and are all non-zero.
This shift of all negative helicity legs in
$M^{\small {\rm N}^2{\rm MHV}}$ allows us to factorise the full amplitude as,
$$
M^{\small {\rm N}^2{\rm MHV}}=\sum_{\alpha}
M^{\small {\rm MHV}}(z_{1,\alpha})\frac{i}{P_{\alpha}^2}
M^{\small {\rm NMHV}}(z_{1,\alpha})\,.
\equn\label{alphasum}
$$
The summation is over all the physical factorisations of the amplitude.
In the above the individual tree amplitudes
are evaluated at the shifted momentum values.
In particular the trees depend upon the shifted, on-shell,
momenta $\hat P_{\alpha,(1)}$.
We consider a single term in the summation corresponding to a specific pole,
$$
D_{\alpha}= M^{\small {\rm MHV}}(z_{1,\alpha})\frac{i}{P_{\alpha}^2}
M^{\small {\rm NMHV}}(z_{1,\alpha})\,,
\equn
$$
and evaluate this by determining the poles in $D_{\alpha}(z_2)$
under the shift,
$$
\bar\lambda_i \longrightarrow
\bar\lambda_i+z_2 r^{(2)}_i\bar\eta , \;\;\;  i=1,\cdots,4 \, .
\equn
$$
The $r^{(2)}_i$ are restricted to maintain momentum conservation
and to leave the pole unshifted,
$$
P^2_\alpha \longrightarrow P^2_\alpha \, ,
\equn
$$
which corresponds to the constraint,
$$
0= \sum_{i}z_2 r^{(2)}_i [ \eta | P_{\alpha} | i \ra \, .
\equn
$$
where $i$ runs over the indices $(1,\ldots,4)$ which lie in the set $\alpha$. 
This condition also implies that the internal legs remain on-shell.

This gives three linear constraints on the four $z_2 r^{(2)}_i$.  The function
$D_{\alpha}(z_2)$ is rational and, since the two tree amplitudes do not have
simultaneous poles, has simple poles. The poles occur where
$M^{\rm NMHV}$ factorises into pairs of MHV amplitudes and thus,  assuming
$D_{\alpha}(z_2)$ vanishes at infinity, we have,
$$
D_{\alpha}=\sum_{\beta}D_{\alpha,\beta}
=\sum_{\beta}  M^{\small{\rm MHV}}(z)\frac{i}{P_{\alpha}^2}
\times
\left(
M^{\small {\rm MHV}} (z)\;
 \frac{i}{P_{\beta,(1)}^2 }
M^{\small{\rm MHV}} (z)\,
\right) \, .
\equn\label{lateadditiuon}
$$
where $z$ indicates the functional dependance upon the two shifts 
(4.10) and (4.13).   
Explicitly, 
all three MHV amplitudes are evaluated at the shifted points,
$$
\bar\lambda_i \to \bar\lambda_i+(z_{1,\alpha}r_i^{(1)}+z_{2,\beta}r_i^{(2)})\bar\eta
.
\equn
$$ In eq.~(\ref{lateadditiuon}) we have a shifted propagator $P_{\beta,(1)}^2$ rather than $P_{\beta}^2$
since we are factorising the shifted tree amplitude $M^{\rm
NMHV}$. Hence this is not immediatelly an MHV diagram term. but is one
contribution to the MHV diagram with unshifted propagators
$i/P_{\alpha}^2$ and $i/P_{\beta}^2$.  There is a second contribution
to the same MHV diagram which arises from the term with a
$P^2_{\beta}$ pole in the sum in eq.~(\ref{alphasum}). Expansion of
this yields,
$$
 D_{\beta,\alpha} =\left(
 M^{\small \rm MHV}(z')\frac{i}{P_{\alpha,(1)}^2}
  M^{\small\rm MHV} (z')\;
\right)
\times \frac{i}{P_{\beta}^2 }
M^{\small\rm MHV} (z')\, ,
\equn
$$
where the MHV amplitudes are now evaluated at the points
$$
\bar\lambda_i \to \bar\lambda_i+(z_{1,\beta}r_i^{(1)}+z_{2,\alpha}'r'{}_i^{(2)})\bar\eta
 \; .
\equn
$$

To prove the MHV-diagram expansion we need to show that the sum of the
two terms $D_{\alpha,\beta}+D_{\beta,\alpha}$ gives the correct diagram, {\it i.e.},
$$
\eqalign{
 M^{\small\rm MHV}(z)\frac{i}{P_{\alpha}^2}
&
\left(
M^{\small\rm MHV} (z)\;
\times \frac{i}{P_{\beta,(1)}^2 }
M^{\small\rm MHV} (z)\,
\right)
\cr
& \hskip 2.0 truecm + \left(
 M^{\small\rm MHV}(z')\frac{i}{P_{\alpha,(1)}^2}
  M^{\small\rm MHV} (z')\;
\right)
\times \frac{i}{P_{\beta}^2 }
M^{\small\rm MHV} (z')\,
\cr
& =M^{\rm MHV}(z_{a})\frac{i}{P^2_{\alpha}} M^{\rm MHV}(z_{a})\frac{i}
{P^2_{\beta}} M^{\rm MHV}(z_{a})\,,
\cr}
\equn
$$
with the  $M^{\rm MHV}(z_{a})$ evaluated at the point $z_{a}$ specified by the
rules of the previous section.

\noindent
We need two facts to show this:

$\bullet$ The product of the three tree amplitudes is the same in both cases
and equal to the desired value.  This is equivalent to showing that
$z\equiv z' \equiv z_a$.

$\bullet$ There is an identity involving the product of propagators,
$$
\frac{i}{P^2_{\alpha}}
\frac{i}{P^2_{\beta,(1)}}
+\frac{i}{P^2_{\alpha,(1)}}\frac{i}{P^2_{\beta}}
=\frac{i}{P^2_{\alpha}}\frac{i}{ P^2_{\beta}}\, .
\equn\label{SecondIdentity}
$$

Taking the first fact:  in the final expression for the
$\bar\lambda_{i}$ the net effect of the two shifts is to give a total shift
of the form,
$$
\hat {\bar\lambda}_{i}=
\bar\lambda_{i} +a_i \bar\eta \, .
\equn
$$
The $a_i$ are such that  momentum conservation is satisfied and
$\hat P^2_{\alpha,(2)} =\hat P^2_{\beta,(2)} =0$.
As discussed in the previous section,
these constraints have a unique solution and so the
$\bar\lambda_{i}$ take the same values irrespective of the
order in which we factorise.  The values of the intermediate momenta,
$\hat P$, are determined by momentum conservation which
are precisely the substitutions specified in the substitution rules.


The second fact can be shown in the following way. 
Considering the contour integral of two shifted propagators
$$
\oint {dz\over z} {1 \over P_\alpha^2(z) P_\beta^2(z)} 
\equn
$$
about a contour at infinity. 
Since the $P_\alpha^2(z)$ vanish at infinity
 this integral vanishes and is also equal to the sum of
its residues. 
Examining the residues we obtain,
$$
{1 \over P_\alpha^2}
{1 \over P_\beta^2}
-{1 \over P_\alpha^2(z_\beta)}
{1 \over P_\beta^2}
-{1 \over P_\alpha^2}
{1 \over P_\beta^2(z_\alpha)} =0 \,,
\equn$$
which provides a proof of eq.(\ref{SecondIdentity}). 
 
Thus  the two terms combine to give a single term
which is the MHV-vertex diagram.

\subsection{General Case}

The general case can be deduced by a repeated application of the process used in
the previous section. We give an outline of this here.
Consider a general ${\rm N}^{n}{\rm MHV}$ amplitude and shift all
the negative helicity legs,
$$
\bar\lambda_i \to \bar\lambda_i+z_1 r_i^{(1)} \bar\eta \, ,
\equn
$$
for a generic set of $r_i^{(1)}$. The amplitude can then be written as,
$$
M^n(0)=\sum_{\alpha}M^{n-k_\alpha+1}(...,\hat p)(z_{1,\alpha})
\frac{i}{P_{\alpha}^2} M^{k_\alpha+1}((-\hat p),...)(z_{1,\alpha})\,.
\label{BCFWexpansion}
\equn
$$
We evaluate an individual term in this by imposing a shift with parameter
$z_2$ that does not shift  $P_{\alpha}^2$. We continue in this way until the
we have an amplitude which is a product of MHV amplitudes with propagators,
$$
D_{\alpha_1,\alpha_2,\cdots \alpha_{n}}
=\prod ( M^{\rm MHV}) \times \prod_i{i \over P_{\alpha_i,(i-1)}^2 } \, ,
\equn
$$
where $i/P^2_{\alpha_i,(i-1)}$ denotes the propagator we factorised on in
the $i$-th step.
As before we gather together all terms with the same pole structure and
combine them into a single diagram. This again requires two things: firstly that the
MHV amplitudes are evaluated at the same point irrespective of the order
and secondly that the pole terms sum to yield the product of the
unshifted poles.

For the first step we note that the net effect of the shifts is to apply an
overall shift to the $n+2$ negative helicity legs of the form,
$$
\hat{\bar\lambda}_i =\bar\lambda_i +a_i\bar\eta\, .
\equn
$$
Since momentum conservation is preserved at each step, overall momentum
conservation is guaranteed at the final stage. This is equivalent to
two linear constraints on the $a_i$.  Secondly, the net effect at
their final stage is that all the $\hat P_{\alpha_i}$ are on-shell, $\hat
P_{\alpha_i}^2=0$.  This imposes $n$ further linear constraints and we are
left with a  unique shift.

Summing over the different orderings now gives an expression of the form,
$$
M^n(0){\big |}_{\mbox{\small CSW-diagram}}=
\left(\prod M^{\rm MHV}\right)\left(\sum_\sigma\prod_{i}
\frac{i}{\hat P_{\alpha_{\sigma(i)},(i-1)}^2}\right)\,,
\equn\label{diagramcontr}
$$
where $\sigma$ denote the permutations of the labels $i=1,...,n$. The rather complicated sum in
(\ref{diagramcontr}) simply yields the product of propagators, as can be seen  by
comparing with the Yang-Mills case.

As the total amplitude can be expressed as a sum of terms, each with
a specific pole structure, the ${\rm N}^n{\rm MHV}$ amplitude $M^n(0)$ can be
written in a CSW form,
$$
M^n(0)=\sum_{\mbox{\small CSW-diagram}}
M^n(0){\big |}_{\mbox{\small CSW-diagram}}\,,
\equn$$
with each CSW diagram contributing as
$$
M^n(0){\big |}_{\mbox{\small CSW-diagram}}
=\left(\prod M^{\rm MHV}\right)\prod_i \frac{i}{P_{\alpha_i}^2}\, ,
\equn$$
as given in the rules of the previous section.

\section{Conclusions and Comments}

In this paper we have shown a new way of  obtaining amplitudes for
graviton scattering, using a gravity MHV-vertex formalism that
resembles the CSW formalism for calculating tree amplitudes in
Yang-Mills theory.  Given the assumption that gravity amplitudes are
sufficiently well behaved under a BCFW-style analytic continuation to
complex momenta, we have presented a direct proof of the formalism and
have illustrated its usefulness through concrete examples such as NMHV
amplitudes.
Although we have presented MHV-vertices for external gravitons only we
expect the procedure to extend to other matter types using
supersymmetry to obtain the relevant MHV-vertex~\cite{Nair1988bq,NairGravity}.

Although the existence of the CSW formalism can be motivated by
the duality with a twistor string theory, such a motivation is not so clear for gravity.
The natural candidate string theories contain conformal
supergravity~\cite{BerkovitsWitten} rather than conventional gravity.
Despite this conventional gravity does seem to share features with
Yang-Mills theory such as the existence of a MHV-vertex construction
and the coplanarity~\cite{BeBbDu} of NMHV amplitudes which hint at the existence of
a twistor string dual theory.

\vspace{1.0cm}
\noindent{\bf Acknowledgments}

We thank Zvi Bern for many useful discussions. This research was supported in part by
the PPARC and the EPSRC of the UK.

\vfill\eject

\end{document}